\documentclass[onecolumn,preprintnumbers,amsmath,amssymb]{revtex4}
\usepackage{graphicx}% Include figure files
\usepackage{amssymb,amsmath}
\usepackage{bm,slashed}% bold math
\usepackage{dcolumn}

\begin{document}

\preprint{}

\title{No Drama Quantum Theory? A Review}% Force line breaks with \\

\author{A. Akhmeteli}
\email{akhmeteli@ltasolid.com}
\affiliation{%
LTASolid Inc.\\
10616 Meadowglen Ln. 2708\\
Houston, TX 77042, USA}%


\homepage{http://www.akhmeteli.org}
%\affiliation{

\date{\today}% It is always \today, today,

\begin{abstract}
Schr\"{o}dinger ~\cite{Schroed} noted that the complex charged matter field in the Klein-Gordon equation or in scalar electrodynamics can be made real by a gauge transform, although it is generally believed that complex functions are required to describe charged fields. Surprisingly, this result can be extended to the Dirac equation or spinor electrodynamics: three out of four complex components of the Dirac spinor function can be algebraically eliminated, and the remaining component can be made real by a gauge transform. Therefore, the Dirac equation is generally equivalent to one fourth-order partial differential equation for one real function (Ref.~\cite{Akhmeteli-JMP}). As the Dirac equation is one of the most fundamental equations, these results both belong in textbooks and can be used for development of new efficient methods and algorithms of quantum chemistry.

The matter field can be algebraically eliminated both in scalar electrodynamics (the Klein-Gordon-Maxwell equation) and in spinor electrodynamics (the Dirac-Maxwell electrodynamics) in a certain gauge (for spinor electrodynamics, this is done after introduction of a complex electromagnetic four-potential, which leaves the electromagnetic fields unchanged). The resulting equations describe independent dynamics of the electromagnetic field (they form closed systems of partial differential equations). This result not only permits mathematical simplification, as the number of fields is reduced, but can also be useful for interpretation of quantum theory. For example, in the Bohm (de Broglie-Bohm) interpretation, the electromagnetic field can replace the wave function as the guiding field. It is also shown that for these systems of equations, a generalized Carleman linearization (Carleman embedding) procedure generates systems of linear equations in the Hilbert space, which look like second-quantized theories and are equivalent to the original nonlinear systems on the set of solutions of the latter. Thus, the relevant local realistic models can be embedded into quantum field theories. These models are equivalent to well-established models - scalar electrodynamics and spinor electrodynamics, so they correctly describe a large body of experimental data. Although they may need some modifications to achieve more complete agreement with experiments, they may be of great interest as "no drama quantum theories", as simple (in principle) as classical electrodynamics. Possible issues with the Bell theorem are discussed.
\end{abstract}

%\keywords{Dirac's "new electrodynamics"; Bohm interpretation; locality.}

\maketitle

\section{Introduction}	%) A SECTION HEADING

Is it possible to offer a "no drama" quantum theory? Something as simple (in principle) as classical electrodynamics - a local realistic theory described by a system of partial differential equations in 3+1 dimensions, but reproducing unitary evolution of quantum theory in the configuration space?

Of course, the Bell inequalities cannot be violated in such a theory. This author has little, if anything, new to say about the Bell theorem, and this article is not about the Bell theorem. However, this issue cannot be "swept under the carpet" and will be discussed in Section V using other people's arguments.

The models of this work are based on scalar electrodynamics (the Klein-Gordon-Maxwell electrodynamics) with equations:
\begin{equation}\label{eq:pr7}
(\partial^\mu+ieA^\mu)(\partial_\mu+ieA_\mu)\psi+m^2\psi=0,
\end{equation}
\begin{equation}\label{eq:pr8}
\Box A_\mu-A^\nu_{,\nu\mu}=j_\mu,
\end{equation}
\begin{equation}\label{eq:pr9}
j_\mu=ie(\psi^*\psi_{,\mu}-\psi^*_{,\mu}\psi)-2e^2 A_\mu\psi^*\psi,
\end{equation}
and spinor electrodynamics (the Dirac-Maxwell electrodynamics) with equations:
\begin{equation}\label{eq:pr25}
(i\slashed{\partial}-\slashed{A})\psi=\psi,
\end{equation}
\begin{equation}\label{eq:pr26}
\Box A_\mu-A^\nu_{,\nu\mu}=e^2\bar{\psi}\gamma_\mu\psi,
\end{equation}
where, e.g., $\slashed{A}=A_\mu\gamma^\mu$ (the Feynman slash notation). For the sake of simplicity, a system of units is used where $\hbar=c=m=1$, and the electric charge $e$ is included in $A_\mu$ for spinor electrodynamics ($eA_\mu \rightarrow A_\mu$).
In the chiral representation of $\gamma$-matrices (Ref.~\cite{Itzykson})
\begin{equation}\label{eq:d1}
\gamma^0=\left( \begin{array}{cc}
0 & -I\\
-I & 0 \end{array} \right),\gamma^i=\left( \begin{array}{cc}
0 & \sigma^i \\
-\sigma^i & 0 \end{array} \right),
\end{equation}
where index $i$ runs from 1 to 3, and $\sigma^i$ are the Pauli matrices.

\maketitle

\section{One real function to describe charged matter field}

Schr\"{o}dinger (Ref.~\cite{Schroed}) noted that the complex charged matter field $\psi$ in scalar electrodynamics (equations (\ref{eq:pr7},\ref{eq:pr8},\ref{eq:pr9})) can be made real by a gauge transform (at least locally), although it is generally believed that complex functions are required to describe charged fields, and the equations of motion in the relevant gauge (unitary gauge) for the transformed 4-potential of electromagnetic field $B^{\mu}$ and real matter field $\varphi$ are as follows:
\begin{equation}\label{eq:pr10}
\Box\varphi-(e^2 B^\mu B_\mu-m^2)\varphi=0,
\end{equation}
\begin{equation}\label{eq:pr11}
\Box B_\mu-B^\nu_{,\nu\mu}=j_\mu,
\end{equation}
\begin{equation}\label{eq:pr12}
j_\mu=-2e^2 B_\mu\varphi^2.
\end{equation}

Work ~\cite{Schroed} has another unique feature. While the initial Klein-Gordon equation~(\ref{eq:pr7}) actually contains two equations for the real and imaginary components of the complex  field $\psi$, the relevant equation (\ref{eq:pr10}) for the real field $\varphi$ contains just one equation. What happened to the other equation? The missing equation is equivalent to the current conservation equation, and the latter can be derived from the Maxwell equations (\ref{eq:pr11}), as the divergence of the antisymmetric tensor vanishes.

It turns out, however, that Schr\"{o}dinger's results also hold in the case of spinor electrodynamics. In general, four complex components of the Dirac spinor function cannot be made real by a single gauge transform, but three complex components out of four can be eliminated from the Dirac equation in a general case, yielding a fourth-order partial differential equation for the remaining component, which component can be made real by a gauge transform. The resulting two equations for one real component can be replaced by one equation plus the current conservation equation, and the latter can be derived from the Maxwell equations.

Spinor electrodynamics is a more realistic theory than scalar electrodynamics, so it seems important that the charged field of spinor electrodynamics can also be described by one real function. It is not clear if similar results can be obtained for the Standard Model.

To illustrate the parallels with beautiful but little-known Schr\"{o}dinger's work ~\cite{Schroed}, the Dirac equation is considered in this section as part of spinor electrodynamics (equations (\ref{eq:pr25},\ref{eq:pr26})), although most derivations are valid without any changes for the Dirac equation in electromagnetic field independently of the Maxwell equations.

The Dirac equation (\ref{eq:pr25}) can be written in components as follows:
\begin{eqnarray}\label{eq:d3}
(A^0+A^3)\psi_3+(A^1-\imath A^2)\psi_4+\imath(\psi_{3,3}-\imath\psi_{4,2}+\psi_{4,1}-\psi_{3,0})=\psi_1,
\end{eqnarray}
\begin{eqnarray}\label{eq:d4}
(A^1+\imath A^2)\psi_3+(A^0-A^3)\psi_4-\imath(\psi_{4,3}-\imath\psi_{3,2}-\psi_{3,1}+\psi_{4,0})=\psi_2,
\end{eqnarray}
\begin{eqnarray}\label{eq:d5}
(A^0-A^3)\psi_1-(A^1-\imath A^2)\psi_2-\imath(\psi_{1,3}-\imath\psi_{2,2}+\psi_{2,1}+\psi_{1,0})=\psi_3,
\end{eqnarray}
\begin{eqnarray}\label{eq:d6}
-(A^1+\imath A^2)\psi_1+(A^0+A^3)\psi_2+\imath\psi_{2,3}+\psi_{1,2}-\imath(\psi_{1,1}+\psi_{2,0})=\psi_4.
\end{eqnarray}
Obviously, equations (\ref{eq:d5},\ref{eq:d6}) can be used to express components $\psi_3,\psi_4$ via $\psi_1,\psi_2$ and eliminate them from equations (\ref{eq:d3},\ref{eq:d4}) (cf. Ref.~\cite{Fay}, p.445). The resulting equations for $\psi_1$ and $\psi_2$ are as follows:
\begin{widetext}
\begin{eqnarray}\label{eq:d7}
\nonumber
-\psi_{1,\mu}^{,\mu}+\psi_2(-\imath A^1_{,3}-A^2_{,3}+A^0_{,2}+A^3_{,2}+
\imath(A^0_{,1}+A^3_{,1}+A^1_{,0})+A^2_{,0})+\\
+\psi_1(-1+A^{\mu} A_{\mu}-\imath A^{\mu}_{,\mu}+\imath A^0_{,3}-A^1_{,2}+A^2_{,1}+\imath A^3_{,0})-2\imath A^{\mu}\psi_{1,\mu}=0,
\end{eqnarray}
\begin{eqnarray}\label{eq:d8}
\nonumber
-\psi_{2,\mu}^{,\mu}+\imath\psi_1( A^1_{,3}+\imath A^2_{,3}+\imath A^0_{,2}-\imath A^3_{,2}+
A^0_{,1}-A^3_{,1}+A^1_{,0}+\imath A^2_{,0})+\\
+\psi_2(-1+A^{\mu} A_{\mu}-\imath( A^{\mu}_{,\mu}+A^0_{,3}+\imath A^1_{,2}-\imath A^2_{,1}+ A^3_{,0}))-2\imath A^{\mu}\psi_{2,\mu}=0.
\end{eqnarray}
\end{widetext}

As equation (\ref{eq:d7}) contains $\psi_2$, but not its derivatives, it can be used to express $\psi_2$ via $\psi_1$:
\begin{eqnarray}\label{eq:d8nn1}
\psi_2=-\left(\imath F^1+F^2\right)^{-1}\left(\Box'+\imath F^3\right)\psi_1,
\end{eqnarray}
where $F^i=E^i+\imath H^i$, electric field $E^i$ and magnetic field $H^i$ are defined by the standard formulas
\begin{eqnarray}\label{eq:d8n1}
F^{\mu\nu}=A^{\nu,\mu}-A^{\mu,\nu}=\left( \begin{array}{cccc}
0 & -E^1 & -E^2 & -E^3\\
E^1 & 0 & -H^3 & H^2\\
E^2 & H^3 & 0 & -H^1\\
E^3 &-H^2 & H^1 & 0  \end{array} \right),
\end{eqnarray}
and the modified d'Alembertian $\Box'$ is defined as follows:
\begin{eqnarray}\label{eq:d8n2}
\Box'=\partial^\mu\partial_\mu+2\imath A^\mu\partial_\mu+\imath A^\mu_{,\mu}-A^\mu A_\mu+1.
\end{eqnarray}
Using the above notation, equation (\ref{eq:d8}) can be rewritten as follows:
\begin{eqnarray}\label{eq:d8nn2}
-\left(\Box'-\imath F^3\right)\psi_2-\left(\imath F^1-F^2\right)\psi_1=0,
\end{eqnarray}
so equation(\ref{eq:d8nn1}) can be used to eliminate $\psi_2$ from equation (\ref{eq:d8nn2}), yielding an equation of the fourth order for $\psi_1$:
\begin{eqnarray}\label{eq:d8n1k}
\left(\left(\Box'-\imath F^3\right)\left(\imath F^1+F^2\right)^{-1}\left(\Box'+\imath F^3\right)-\imath F^1+F^2\right)\psi_1=0.
\end{eqnarray}

This equation is equivalent to the Dirac equation (if $\imath F^1+F^2\slashed{\equiv}0$).

It should be noted that the coefficient at $\psi_2$ in equation (\ref{eq:d7}) is gauge-invariant (it can be expressed via electromagnetic fields). While this elimination could not be performed for zero electromagnetic fields, this does not look like a serious limitation, as in reality there always exist electromagnetic fields in the presence of charged fields, although they may be very small. However, it should be noted that free spinor field presents a special case and is not considered in this work, as it does not satisfy the equations of spinor electrodynamics. It is not clear how free field being a special case is related to the divergencies in quantum electrodynamics. It should also be noted that the above procedure could be applied to any component of the spinor function, not just to $\psi_1$, yielding equations similar to equation (\ref{eq:d8n1k}). Presenting the above results in a more symmetric form is beyond the scope of this work.

While the above elimination of the third component of the Dirac spinor is straightforward, this author failed to find it elsewhere, but cannot be sure that this important result was not published previously.

Using a gauge transform, it is possible to make $\psi_1$ real (at least locally). Then the real and the imaginary parts of equation (\ref{eq:d8}) after substitution of the expression for $\psi_2$ will present two equations for $\psi_1$. However, it is possible to construct just one equation for $\psi_1$ in such a way that the system containing this equation and the current conservation equation will be equivalent to equation (\ref{eq:d8}).

Let us consider the current conservation equation:
\begin{equation}\label{eq:d9}
(\bar{\psi}\gamma^\mu\psi)_{,\mu}=0,
\end{equation}
or
\begin{equation}\label{eq:d10}
(\bar{\psi}_{,\mu}\gamma^\mu\psi)+(\bar{\psi}\gamma^\mu\psi_{,\mu})=0,
\end{equation}
On the other hand,
\begin{eqnarray}\label{eq:d11}
(\bar{\psi}_{,\mu}\gamma^\mu\psi)^*=(\bar{\psi}_{,\mu}\gamma^\mu\psi)^\dag=\psi^\dag(\gamma^\mu)^\dag(\psi^\dag_{,\mu}\gamma^0)^\dag=
\bar{\psi}\gamma^0\gamma^0\gamma^\mu\gamma^0\gamma^0\psi_{,\mu}=\bar{\psi}\gamma^\mu\psi_{,\mu},
\end{eqnarray}
as $(\gamma^\mu)^\dag=\gamma^0\gamma^\mu\gamma^0$ (Ref.~\cite{Itzykson}), so the current conservation equation can be written as follows:
\begin{eqnarray}\label{eq:d12}
2\textrm{Re}(\bar{\psi}\gamma^\mu\psi_{,\mu})=2\textrm{Im}(\bar{\psi}\imath\slashed{\partial}\psi)=2\textrm{Im}(\bar{\psi}(\imath\slashed{\partial}-\slashed{A}-1)\psi)=0,
\end{eqnarray}
as it is not difficult to check that values $\bar{\psi}\slashed{A}\psi$ and $\bar{\psi}\psi$ are real.

If equations (\ref{eq:d5},\ref{eq:d6},\ref{eq:d7}) hold, only the second component of spinor $(\imath\slashed{\partial}-\slashed{A}-1)\psi$ can be nonzero, so equation (\ref{eq:d12}) can be written as follows:
\begin{equation}\label{eq:d13}
2\textrm{Im}(-\psi_4^* \delta)=0,
\end{equation}
where $\delta$ is the left-hand side of equation (\ref{eq:d8}). Therefore, if $\psi_4$ does not vanish identically, the system containing the current conservation equation equation (\ref{eq:d13}) and the following equation
\begin{equation}\label{eq:d14}
2\textrm{Re}(\psi_4^* \delta)=0
\end{equation}
is equivalent to equation (\ref{eq:d8}).

\section{Elimination of Matter Field and Independent Evolution of Electromagnetic Field}

The following unexpected result was proven in Ref.~\cite{Akhm10} (see also Refs.~\cite{Akhmeteli-IJQI,Akhm18}): the equations obtained from equations ~(\ref{eq:pr10},\ref{eq:pr11},\ref{eq:pr12}) after natural elimination of the matter field form a closed system of partial differential equations and thus describe independent dynamics of electromagnetic field. The detailed wording is as follows: if components of the 4-potential of the electromagnetic field and their first derivatives with respect to time are known in the entire space at some time point, the values of their second derivatives with respect to time can be calculated for the same time point, so the Cauchy problem can be posed, and integration yields the 4-potential in the entire space-time. Thus, the broad range of quantum phenomena described by the scalar electrodynamics can be described in terms of electromagnetic field only. This result not only permits mathematical simplification, as the number of fields is reduced, but can also be useful for interpretation of quantum theory. For example, in the Bohm (de Broglie-Bohm) interpretation (Refs.~\cite{BohmHiley,Holland,Goldstein}), the electromagnetic field can replace the wave function as the guiding field. This may make the interpretation more attractive, removing, for example, the reason for the following criticism of the Bohm interpretation: "If one believes that the particles are real one must also believe the wavefunction is real because it determines the actual trajectories of the particles. This allows us to have a realist interpretation which solves the measurement problem, but the cost is to believe in a double ontology.~\cite{Smolin}" Independent of the interpretation, quantum phenomena can be described in terms of electromagnetic field only.

To eliminate the matter field $\varphi$ from Eqs.~(\ref{eq:pr10},\ref{eq:pr11},\ref{eq:pr12}), let us use a substitution $\Phi=\varphi^2$ first. For example, as
\begin{equation}\label{eq:pr1q}
\Phi_{,\mu}=2\varphi\varphi_{,\mu},
\end{equation}
we obtain
\begin{equation}\label{eq:pr2q}
\Phi_{,\mu}^{,\mu}=2\varphi^{,\mu}\varphi_{,\mu}+2\varphi\varphi^{,\mu}_{,\mu}=
\frac{1}{2}\frac{\Phi^{,\mu}\Phi_{,\mu}}{\Phi}+2\varphi\varphi^{,\mu}_{,\mu}.
\end{equation}
Multiplying Eq.~(\ref{eq:pr10}) by $2\varphi$, we obtain the following equations in terms of $\Phi$ instead of Eqs.~(\ref{eq:pr10},\ref{eq:pr11},\ref{eq:pr12}):
\begin{equation}\label{eq:pr3q}
\Box\Phi-\frac{1}{2}\frac{\Phi^{,\mu}\Phi_{,\mu}}{\Phi}-2(e^2 B^\mu B_\mu-m^2)\Phi=0,
\end{equation}
\begin{equation}\label{eq:pr4q}
\Box B_\mu-B^\nu_{,\nu\mu}=-2e^2 B_\mu\Phi,
\end{equation}
To prove that these equations describe independent evolution of the electromagnetic field $B^\mu$, it is sufficient to prove that if components $B^\mu$ of the potential and their first derivatives with respect to $x^0$ ($\dot{B}^\mu$) are known in the entire space at some time point $x^0=\rm{const}$  (that means that all spatial derivatives of these values are also known in the entire space at that time point), equations ~(\ref{eq:pr3q},\ref{eq:pr4q}) yield the values of their second derivatives, $\ddot{B}^\mu$, for the same value of $x^0$. Indeed, $\Phi$ can be eliminated using equation ~(\ref{eq:pr4q}) for $\mu=0$, as this equation does not contain $\ddot{B}^\mu$ for this value of $\mu$:
\begin{equation}\label{eq:pr5q}
\Phi=(-2e^2 B_0)^{-1}(\Box B_0-B^\nu_{,\nu 0})=(-2e^2 B_0)^{-1}(B^{,i}_{0,i}-B^i_{,i 0})
\end{equation}
(Greek indices in the Einstein sum convention run from $0$ to $3$, and Latin indices run from $1$ to $3$).
Then $\ddot{B}^i$ ($i=1,2,3$) can be determined by substitution of equation ~(\ref{eq:pr5q}) into equation ~(\ref{eq:pr4q}) for $\mu=1,2,3$:
\begin{equation}\label{eq:pr6q}
\ddot{B}_i=-B^{,j}_{i,j}+B^\nu_{,\nu i}+(B_0)^{-1} B_i(B^{,j}_{0,j}-B^j_{,j 0}).
\end{equation}
Thus, to complete the proof, we only need to find $\ddot{B}^0$. Conservation of current implies
\begin{equation}\label{eq:pr7q}
0=(B^\mu \Phi)_{,\mu}=B^\mu_{,\mu}\Phi+B^\mu\Phi_{,\mu}.
\end{equation}
This equation determines $\dot{\Phi}$, as spatial derivatives of $\Phi$ can be found from equation ~(\ref{eq:pr5q}). Differentiation of this equation with respect to $x^0$ yields
\begin{eqnarray}\label{eq:pr8q}
0=(\ddot{B}^0+\dot{B}^i_{,i})\Phi+(\dot{B}^0+B^i_{,i})\dot{\Phi}+\dot{B}^0\dot{\Phi}+\dot{B}^i\Phi_{,i}+B^0\ddot{\Phi}+B^i\dot{\Phi}_{,i}.
\end{eqnarray}
After substitution of $\Phi$ from equation ~(\ref{eq:pr5q}), $\dot{\Phi}$ from equation ~(\ref{eq:pr7q}), and $\ddot{\Phi}$ from  equation ~(\ref{eq:pr3q}) into  equation ~(\ref{eq:pr8q}), the latter equation determines $\ddot{B^0}$ as a function of $B^\mu$, $\dot{B}^\mu$ and their spatial derivatives (again, spatial derivatives of $\Phi$ and $\dot{\Phi}$ can be found from the expressions for $\Phi$ and $\dot{\Phi}$ as functions of $B^\mu$ and $\dot{B}^\mu$). Thus, if $B^\mu$ and $\dot{B}^\mu$ are known in the entire space at a certain value of $x^0$, then $\ddot{B}^\mu$ can be calculated for the same $x^0$, so integration yields $B^\mu$ in the entire space-time. Therefore, we do have independent dynamics of electromagnetic field.

Similar results can be obtained for spinor electrodynamics (equations (\ref{eq:pr25},\ref{eq:pr26})).

Let us apply the following "generalized gauge transform":
\begin{equation}\label{eq:dd1}
\psi=\exp(i\alpha)\varphi,
\end{equation}
\begin{equation}\label{eq:dd1a}
A_\mu=B_\mu-\alpha_{,\mu},
\end{equation}
where the new four-potential $B_\mu$ is complex, $\alpha=\alpha(x^\mu)=\beta+i\delta$, $\beta=\beta(x^\mu)$, $\delta=\delta(x^\mu)$, and $\beta$, $\delta$ are real. The imaginary part of the complex four-potential is a gradient of a certain function, so alternatively we can use this function instead of the imaginary components of the four-potential.

After the transform, the equations of spinor electrodynamics can be rewritten as follows:
\begin{equation}\label{eq:dd2}
(i\slashed{\partial}-\slashed{B})\varphi=\varphi,
\end{equation}
\begin{equation}\label{eq:dd3}
\Box B_\mu-B^\nu_{,\nu\mu}=\exp(-2\delta)e^2\bar{\varphi}\gamma_\mu\varphi.
\end{equation}
If $\psi$ and $\varphi$ have components
\begin{equation}\label{eq:dd4}
\varphi=\left( \begin{array}{c}
\varphi_1\\
\varphi_2\\
\varphi_3\\
\varphi_4\end{array}\right),
\psi=\left( \begin{array}{c}
\psi_1\\
\psi_2\\
\psi_3\\
\psi_4\end{array}\right),
\end{equation}
let us fix the "gauge transform" of equations (\ref{eq:dd1},\ref{eq:dd1a}) somewhat arbitrarily by the following condition:
\begin{equation}\label{eq:dd5}
\varphi_1=\exp(-i\alpha)\psi=1.
\end{equation}
The Dirac equation (\ref{eq:dd2}) can be written in components as follows:
\begin{eqnarray}\label{eq:dd6}
(B^0+B^3)\varphi_3+(B^1-i B^2)\varphi_4+i(\varphi_{3,3}-i\varphi_{4,2}+\varphi_{4,1}-\varphi_{3,0})=\varphi_1,
\end{eqnarray}
\begin{eqnarray}\label{eq:dd7}
(B^1+i B^2)\varphi_3+(B^0-B^3)\varphi_4-i(\varphi_{4,3}-i\varphi_{3,2}-\varphi_{3,1}+\varphi_{4,0})=\varphi_2,
\end{eqnarray}
\begin{eqnarray}\label{eq:dd8}
(B^0-B^3)\varphi_1-(B^1-i B^2)\varphi_2-i(\varphi_{1,3}-i\varphi_{2,2}+\varphi_{2,1}+\varphi_{1,0})=\varphi_3,
\end{eqnarray}
\begin{eqnarray}\label{eq:dd9}
-(B^1+i B^2)\varphi_1+(B^0+B^3)\varphi_2+i\varphi_{2,3}+\varphi_{1,2}-i(\varphi_{1,1}+\varphi_{2,0})=\varphi_4.
\end{eqnarray}
Equations (\ref{eq:dd8},\ref{eq:dd9}) can be used to express components $\varphi_3,\varphi_4$ via $\varphi_1,\varphi_2$ and eliminate them from equations (\ref{eq:dd6},\ref{eq:dd7}). The resulting equations for $\varphi_1$ and $\varphi_2$ are as follows:
\begin{widetext}
\begin{eqnarray}\label{eq:dd10}
\nonumber
-\varphi_{1,\mu}^{,\mu}+\varphi_2(-i B^1_{,3}-B^2_{,3}+B^0_{,2}+B^3_{,2}+
i(B^0_{,1}+B^3_{,1}+B^1_{,0})+B^2_{,0})+\\
+\varphi_1(-1+B^{\mu} B_{\mu}-i B^{\mu}_{,\mu}+i B^0_{,3}-B^1_{,2}+B^2_{,1}+i B^3_{,0})-2i B^{\mu}\varphi_{1,\mu}=0,
\end{eqnarray}
\begin{eqnarray}\label{eq:dd11}
\nonumber
-\varphi_{2,\mu}^{,\mu}+i\varphi_1( B^1_{,3}+i B^2_{,3}+i B^0_{,2}-i B^3_{,2}+
B^0_{,1}-B^3_{,1}+B^1_{,0}+i B^2_{,0})+\\
+\varphi_2(-1+B^{\mu} B_{\mu}-i( B^{\mu}_{,\mu}+B^0_{,3}+i B^1_{,2}-i B^2_{,1}+ B^3_{,0}))-2i B^{\mu}\varphi_{2,\mu}=0.
\end{eqnarray}
\end{widetext}

Equation (\ref{eq:dd10}) can be used to express $\varphi_2$ via $\varphi_1$:
\begin{eqnarray}\label{eq:dd12}
\varphi_2=-\left(i F^1+F^2\right)^{-1}\left(\Box'+i F^3\right)\varphi_1,
\end{eqnarray}
where $F^i=E^i+i H^i$, real electric field $E^i$ and magnetic field $H^i$ are defined by the standard formulas
\begin{eqnarray}\label{eq:dd13}
F^{\mu\nu}=B^{\nu,\mu}-B^{\mu,\nu}=\left( \begin{array}{cccc}
0 & -E^1 & -E^2 & -E^3\\
E^1 & 0 & -H^3 & H^2\\
E^2 & H^3 & 0 & -H^1\\
E^3 &-H^2 & H^1 & 0  \end{array} \right),
\end{eqnarray}
and the modified d'Alembertian $\Box'$ is defined as follows:
\begin{eqnarray}\label{eq:dd14}
\Box'=\partial^\mu\partial_\mu+2 i B^\mu\partial_\mu+i B^\mu_{,\mu}-B^\mu B_\mu+1.
\end{eqnarray}
Equation (\ref{eq:dd11}) can be rewritten as follows:
\begin{eqnarray}\label{eq:dd15}
-\left(\Box'-i F^3\right)\varphi_2-\left(i F^1-F^2\right)\varphi_1=0.
\end{eqnarray}
Substitution of $\varphi_2$ from equation (\ref{eq:dd12}) into equation (\ref{eq:dd11}) yields an equation of the fourth order for $\varphi_1$:
\begin{eqnarray}\label{eq:dd16}
\left(\left(\Box'-i F^3\right)\left(i F^1+F^2\right)^{-1}\left(\Box'+i F^3\right)-i F^1+F^2\right)\varphi_1=0.
\end{eqnarray}

Application of the gauge condition of equation (\ref{eq:dd5}) to equations (\ref{eq:dd14},\ref{eq:dd12},\ref{eq:dd16}, and \ref{eq:dd15}) yields the following equations:
\begin{eqnarray}\label{eq:dd17}
\Box'\varphi_1=i B^\mu_{,\mu}-B^\mu B_\mu+1,
\end{eqnarray}
\begin{eqnarray}\label{eq:dd18}
\varphi_2=-\left(i F^1+F^2\right)^{-1}\left(i B^\mu_{,\mu}-B^\mu B_\mu+1+i F^3\right),
\end{eqnarray}
\begin{eqnarray}\label{eq:dd19}
\left(\Box'-i F^3\right)\left(i F^1+F^2\right)^{-1}\left(i B^\mu_{,\mu}-B^\mu B_\mu+1+i F^3\right)-i F^1+F^2=0,
\end{eqnarray}
\begin{eqnarray}\label{eq:dd20b}
-\left(\Box'-i F^3\right)\varphi_2-\left(i F^1-F^2\right)=0.
\end{eqnarray}
Obviously, equations (\ref{eq:dd5},\ref{eq:dd18},\ref{eq:dd8}, and \ref{eq:dd9}) can be used to eliminate spinor $\varphi$ from the equations of spinor electrodynamics (\ref{eq:dd2},\ref{eq:dd3}). It is then possible to eliminate $\delta$ from the resulting equations. Furthermore, it turns out that the equations describe independent dynamics of the (complex four-potential of) electromagnetic field $B^\mu$. More precisely, if components $B^\mu$ and their temporal derivatives (derivatives with respect to $x^0$) up to the second order $\dot{B}^\mu$ and $\ddot{B}^\mu$ are known at some point in time in the entire 3D space $x^0$=const, the equations determine the temporal derivatives of the third order $\dddot{B}^\mu$, so the Cauchy problem can be posed, and the equations can be integrated (at least locally). Let us prove this statement.

As $\varphi_1$=1 (equation (\ref{eq:dd5})), we obtain
\begin{eqnarray}\label{eq:dd20}
\bar{\varphi}\gamma_\mu\varphi=\left( \begin{array}{c}
\varphi_1^*\varphi_1+\varphi_2^*\varphi_2+\varphi_3^*\varphi_3+\varphi_4^*\varphi_4\\
\varphi_2^*\varphi_1+\varphi_1^*\varphi_2-\varphi_4^*\varphi_3-\varphi_3^*\varphi_4\\
i\varphi_2^*\varphi_1-i\varphi_1^*\varphi_2-i\varphi_4^*\varphi_3+i\varphi_3^*\varphi_4\\
\varphi_1^*\varphi_1-\varphi_2^*\varphi_2-\varphi_3^*\varphi_3+\varphi_4^*\varphi_4\end{array}\right)=
\left( \begin{array}{c}
1+\varphi_2^*\varphi_2+\varphi_3^*\varphi_3+\varphi_4^*\varphi_4\\
\varphi_2^*+\varphi_2-\varphi_4^*\varphi_3-\varphi_3^*\varphi_4\\
i\varphi_2^*-i\varphi_2-i\varphi_4^*\varphi_3+i\varphi_3^*\varphi_4\\
1-\varphi_2^*\varphi_2-\varphi_3^*\varphi_3+\varphi_4^*\varphi_4\end{array}\right).
\end{eqnarray}
Using equation (\ref{eq:dd3}) with index $\mu=0$ and equation (\ref{eq:dd20}), we can express $e^2\exp(-2\delta)$ as follows:
\begin{eqnarray}\label{eq:dd21}
e^2\exp(-2\delta)=\left(B^{,i}_{0,i}-B^i_{,i0}\right)
(1+\varphi_2^*\varphi_2+\varphi_3^*\varphi_3+\varphi_4^*\varphi_4)^{-1},
\end{eqnarray}
as
\begin{eqnarray}\label{eq:dd22a}
\Box B_0-B^\nu_{,\nu 0}=B^{,i}_{0,i}-B^i_{,i0}
\end{eqnarray}
(Latin indices run from 1 to 3, and Greek indices run from 0 to 3). Substitution of equation (\ref{eq:dd21}) in equation (\ref{eq:dd3}) yields
\begin{eqnarray}\label{eq:dd23a}
\nonumber
\Box B_i-B^\nu_{,\nu i}=\ddot{B}_i+B_{i,j}^{,j}-\dot{B}^0_{,i}-B^j_{,j i}=\\
\left(B^{,j}_{0,j}-B^j_{,j0}\right)(1+\varphi_2^*\varphi_2+\varphi_3^*\varphi_3+\varphi_4^*\varphi_4)^{-1}
\left(\begin{array}{c}
\varphi_2^*+\varphi_2-\varphi_4^*\varphi_3-\varphi_3^*\varphi_4\\
i\varphi_2^*-i\varphi_2-i\varphi_4^*\varphi_3+i\varphi_3^*\varphi_4\\
1-\varphi_2^*\varphi_2-\varphi_3^*\varphi_3+\varphi_4^*\varphi_4\end{array}\right).
\end{eqnarray}
We note based on equation (\ref{eq:dd18}) that $\varphi_2$ can be expressed via $B^\mu$, $\dot{B}^\mu$, and their spatial derivatives (derivatives with respect to $x^1$, $x^2$, and $x^3$), as
\begin{equation}\label{eq:dd22}
F^1=E^1+i H^1=F^{10}+i F^{32}=B^{0,1}-B^{1,0}+i(B^{2,3}-B^{3,2}),
\end{equation}
\begin{equation}\label{eq:dd22b}
F^2=E^2+i H^2=F^{20}+i F^{13}=B^{0,2}-B^{2,0}+i(B^{3,1}-B^{1,3}),
\end{equation}
\begin{equation}\label{eq:dd22c}
F^3=E^3+i H^3=F^{30}+i F^{21}=B^{0,3}-B^{3,0}+i(B^{1,2}-B^{2,1}).
\end{equation}
Using equations (\ref{eq:dd22},\ref{eq:dd22b},\ref{eq:dd22c}), the first temporal derivatives of $F^i$ can be written as follows:
\begin{equation}\label{eq:dd23}
\dot{F}^1=\dot{B}^{0,1}-\ddot{B}^{1}+i(\dot{B}^{2,3}-\dot{B}^{3,2}),
\end{equation}
\begin{equation}\label{eq:dd23b}
\dot{F}^2=\dot{B}^{0,2}-\ddot{B}^{2}+i(\dot{B}^{3,1}-\dot{B}^{1,3}),
\end{equation}
\begin{equation}\label{eq:dd23c}
\dot{F}^3=\dot{B}^{0,3}-\ddot{B}^{3}+i(\dot{B}^{1,2}-\dot{B}^{2,1}).
\end{equation}
We note based on equations (\ref{eq:dd18},\ref{eq:dd22},\ref{eq:dd22b},\ref{eq:dd22c},\ref{eq:dd23},\ref{eq:dd23b},\ref{eq:dd23c}) that $\dot{\varphi}_2$ can be expressed via $B^\mu$, $\dot{B}^\mu$, $\ddot{B}^\mu$, and their spatial derivatives.

From equations (\ref{eq:dd8},\ref{eq:dd5}) we obtain:
\begin{eqnarray}\label{eq:dd24}
\varphi_3=B^0-B^3-(B^1-i B^2)\varphi_2-i(-i\varphi_{2,2}+\varphi_{2,1}).
\end{eqnarray}
We note that that $\varphi_3$ can be expressed via $B^\mu$, $\dot{B}^\mu$, and their spatial derivatives.
The first temporal derivative of $\varphi_3$ can be written as follows:
\begin{eqnarray}\label{eq:dd25}
\dot{\varphi}_3=\dot{B}^0-\dot{B}^3-(\dot{B}^1-i \dot{B}^2)\varphi_2-(B^1-i B^2)\dot{\varphi}_2-i(-i \dot{\varphi}_{2,2}+ \dot{\varphi}_{2,1}).
\end{eqnarray}
We note based on equation (\ref{eq:dd25}) that $\dot{\varphi}_3$ can be expressed via $B^\mu$, $\dot{B}^\mu$, $\ddot{B}^\mu$, and their spatial derivatives.

From equation (\ref{eq:dd9},\ref{eq:dd5}) we obtain:
\begin{eqnarray}\label{eq:dd26}
\varphi_4=-(B^1+i B^2)+(B^0+B^3)\varphi_2+i\varphi_{2,3}-i\varphi_{2,0}.
\end{eqnarray}
We note that $\varphi_4$ can be expressed via $B^\mu$, $\dot{B}^\mu$, $\ddot{B}^\mu$, and their spatial derivatives.
The first temporal derivative of $\varphi_4$ can be written as follows:
\begin{eqnarray}\label{eq:dd26k}
\dot{\varphi}_4=-(\dot{B}^1+i \dot{B}^2)+(\dot{B}^0+\dot{B}^3)\varphi_2+(B^0+B^3)\dot{\varphi_2}+i\dot{\varphi}_{2,3}-i\ddot{\varphi}_2.
\end{eqnarray}
All terms in the expression for $\dot{\varphi}_4$ with a possible exception of $-i\ddot{\varphi}_2$ can be expressed via $B^\mu$, $\dot{B}^\mu$, $\ddot{B}^\mu$, and their spatial derivatives. Let us consider the expression $\ddot{\varphi}_2$.

Equations (\ref{eq:dd20b},\ref{eq:dd14}) yield:
\begin{eqnarray}\label{eq:dd27}
\nonumber
0=-\left(\Box'-i F^3\right)\varphi_2-\left(i F^1-F^2\right)=\\
\nonumber
-\left(\partial^\mu\partial_\mu+2 i B^\mu\partial_\mu+i B^\mu_{,\mu}-B^\mu B_\mu+1-i F^3\right)\varphi_2-\left(i F^1-F^2\right)=\\
\nonumber
-\left(\partial^0\partial_0+\partial^i\partial_i+2 i B^0\partial_0+2 i B^i\partial_i+i B^\mu_{,\mu}-B^\mu B_\mu+1-i F^3\right)\varphi_2-\left(i F^1-F^2\right)=\\
-\ddot{\varphi}_2-2 i B^0\dot{\varphi}_2-\left(\partial^i\partial_i+2 i B^i\partial_i+i B^\mu_{,\mu}-B^\mu B_\mu+1-i F^3\right)\varphi_2-\left(i F^1-F^2\right).
\end{eqnarray}
We note that $\ddot{\varphi}_2$ can be expressed via $B^\mu$, $\dot{B}^\mu$, $\ddot{B}^\mu$, and their spatial derivatives. Therefore, based on equation (\ref{eq:dd26k}), the same is true for $\dot{\varphi}_4$. Furthermore, we can summarize that all functions $\varphi_\mu$ and $\dot{\varphi}_\mu$ can be expressed via $B^\mu$, $\dot{B}^\mu$, $\ddot{B}^\mu$, and their spatial derivatives. Obviously, the same is true for $\varphi_\mu^*$ and $\dot{\varphi}_\mu^*$.

Differentiating equations (\ref{eq:dd23a}) with respect to time ($x^0$), we conclude that functions $\dddot{B}^i$ can be expressed via $B^\mu$, $\dot{B}^\mu$, $\ddot{B}^\mu$, and their spatial derivatives, as the left-hand side of equations (\ref{eq:dd23a}) after the differentiation equals
\begin{eqnarray}\label{eq:dd28}
\dddot{B}_i+\dot{B}_{i,j}^{,j}-\ddot{B}^0_{,i}-\dot{B}^j_{,j i},
\end{eqnarray}
and the right-hand side of equation (\ref{eq:dd23a}) after the differentiation will be expressed via $B^\mu$, $\dot{B}^\mu$, $\ddot{B}^\mu$, $\varphi_\mu$, $\dot{\varphi}_\mu$, $\varphi_\mu^*$, $\dot{\varphi}_\mu^*$, and their spatial derivatives. Therefore, functions $\dddot{B}_i$ can be expressed via $B^\mu$, $\dot{B}^\mu$, $\ddot{B}^\mu$, and their spatial derivatives, so to prove the initial statement we just need to prove the same for $\dddot{B}_0$. To this end, let us consider the following equation derived from equations (\ref{eq:dd19},\ref{eq:dd14}):
\begin{eqnarray}\label{eq:dd29}
\left(\partial^\mu\partial_\mu+2 i B^\mu\partial_\mu+i B^\mu_{,\mu}-B^\mu B_\mu+1-i F^3\right)\left(i
F^1+F^2\right)^{-1}\left(i B^\mu_{,\mu}-B^\mu B_\mu+1+i F^3\right)-
i F^1+F^2=0.
\end{eqnarray}
It is obvious that the following part of the left-hand side of equation (\ref{eq:dd29}) can be expressed via $B^\mu$, $\dot{B}^\mu$, and their spatial derivatives:
\begin{eqnarray}\label{eq:dd30}
\left(i B^\mu_{,\mu}-B^\mu B_\mu+1-i F^3\right)\left(i
F^1+F^2\right)^{-1}\left(i B^\mu_{,\mu}-B^\mu B_\mu+1+i F^3\right)-
i F^1+F^2.
\end{eqnarray}
The rest of the left-hand side of equation (\ref{eq:dd29}) can be rewritten as follows:
\begin{eqnarray}\label{eq:dd31}
\left(\partial^0\partial_0+\partial^i\partial_i+2 i B^0\partial_0+2 i B^i\partial_i\right)\left(i
F^1+F^2\right)^{-1}\left(i B^\mu_{,\mu}-B^\mu B_\mu+1+i F^3\right).
\end{eqnarray}
The following part of the expression in equation (\ref{eq:dd31}) can be expressed via $B^\mu$, $\dot{B}^\mu$, and their spatial derivatives:
\begin{eqnarray}\label{eq:dd31a}
\left(\partial^i\partial_i+2 i B^i\partial_i\right)\left(i
F^1+F^2\right)^{-1}\left(i B^\mu_{,\mu}-B^\mu B_\mu+1+i F^3\right).
\end{eqnarray}
Let us evaluate the following expression:
\begin{eqnarray}\label{eq:dd32}
\nonumber
\partial_0\left(i F^1+F^2\right)^{-1}\left(i B^\mu_{,\mu}-B^\mu B_\mu+1+i F^3\right)=\\
-\left(i \dot{F}^1+\dot{F}^2\right)\left(i F^1+F^2\right)^{-2}\left(i B^\mu_{,\mu}-B^\mu B_\mu+1+i F^3\right)+
\left(i F^1+F^2\right)^{-1}\left(i \dot{B}^\mu_{,\mu}-2\dot{B}^\mu B_\mu+i \dot{F}^3\right).
\end{eqnarray}
Thus, the term $2 i B^0\partial_0$ in the first pair of parentheses of Eq.(\ref{eq:dd31}) produces terms that can be expressed via $B^\mu$, $\dot{B}^\mu$, $\ddot{B}^\mu$, and their spatial derivatives. Therefore, we only need to evaluate (using equation (\ref{eq:dd32})) the following expression:
\begin{eqnarray}\label{eq:dd33}
\nonumber
\partial^0\partial_0\left(i F^1+F^2\right)^{-1}\left(i B^\mu_{,\mu}-B^\mu B_\mu+1+i F^3\right)=\\
\nonumber
\partial^0\left(-\left(i \dot{F}^1+\dot{F}^2\right)\left(i F^1+F^2\right)^{-2}\left(i B^\mu_{,\mu}-B^\mu B_\mu+1+i F^3\right)\right)+
\partial^0\left(i F^1+F^2\right)^{-1}\left(i \dot{B}^\mu_{,\mu}-2\dot{B}^\mu B_\mu+i \dot{F}^3\right)=\\
\nonumber
\partial^0\left(-\left(i \dot{F}^1+\dot{F}^2\right)\left(i F^1+F^2\right)^{-2}\left(i B^\mu_{,\mu}-B^\mu B_\mu+1+i F^3\right)\right)+
\left(\partial^0\left(i F^1+F^2\right)^{-1}\right)\left(i \dot{B}^\mu_{,\mu}-2\dot{B}^\mu B_\mu+i \dot{F}^3\right)+\\
\left(i F^1+F^2\right)^{-1}\left(i \dddot{B}^0+i \ddot{B}^i_{,i}+\left(\partial^0\left(-2\dot{B}^\mu B_\mu+i \dot{F}^3\right)\right)\right).
\end{eqnarray}
It follows from equations (\ref{eq:dd23},\ref{eq:dd23b},\ref{eq:dd23c}) that $\ddot{F^i}$ can be expressed via $B^\mu$, $\dot{B}^\mu$, $\ddot{B}^\mu$, $\dddot{B}^i$ (but not $\dddot{B}^0$), and their spatial derivatives, but, as explained above, $\dddot{B}^i$ can be expressed via $B^\mu$, $\dot{B}^\mu$, $\ddot{B}^\mu$, and their spatial derivatives. Thus, this is also true for all terms of equation (\ref{eq:dd33}) (and, consequently, equation (\ref{eq:dd29})), with a possible exception of the term
\begin{eqnarray}\label{eq:dd33a}
\left(i F^1+F^2\right)^{-1}\dddot{B^0},
\end{eqnarray}
but that means that equation (\ref{eq:dd29}) can be used to express $\dddot{B^0}$ via $B^\mu$, $\dot{B}^\mu$, $\ddot{B}^\mu$, and their spatial derivatives, which completes the proof.

\section{Transition to Many-Particle Theories}
The theories we considered in the previous section are not second-quantized, so, on the face of it, they cannot describe many particles. On the other hand, nightlight (Ref.~\cite{nightlight2}) indicated that, rather amazingly, second-quantized theories (or at least theories that look like second-quantized ones) can be obtained from nonlinear partial differential equations by a generalization of the Carleman linearization (Carleman embedding) procedure (Ref.~\cite{Kowalski}). This generalized procedure generates for a system of nonlinear partial differential equations a system of linear equations in the Hilbert space, which looks like a second-quantized theory and is equivalent to the original nonlinear system on the set of solutions of the latter.

Following Ref.~\cite{Kowalski}, let us consider a nonlinear differential equation in an (s+1)-dimensional space-time (the equations describing independent dynamics of electromagnetic field for scalar electrodynamics and spinor electrodynamics are a special case of this equation) ${\partial_t}\boldsymbol{\xi}(x,t) = \boldsymbol{F}(\boldsymbol{\xi},{D^\alpha}\boldsymbol{\xi};x,t)$  , $\boldsymbol{\xi}(x,0)=\boldsymbol{\xi}_0(x)$, where $\boldsymbol{\xi}:\mathbf{R}^s\times\mathbf{R}\rightarrow\mathbf{C}^k$, $D^\alpha\boldsymbol{\xi}=\left(D^{\alpha_1}\xi_1,\ldots ,D^{\alpha_k}\xi_k\right)$, $\alpha_i$ are multiindices, ${D^\beta}={\partial^{|\beta|}}/\partial x_1^{\beta_1}\ldots\partial x_s^{\beta_s}$, with $ |\beta|=\sum\limits_{i=1}^{s}\beta_i$, is a generalized derivative, $\boldsymbol{F}$ is analytic in $\boldsymbol{\xi}$, $D^\alpha\boldsymbol{\xi}$. It is also assumed that $\boldsymbol{\xi_0}$ and $\boldsymbol{\xi}$ are square integrable. Then Bose operators $\boldsymbol{a^\dag(x)}=\left(a^\dag_1(x),\ldots,a^\dag_k(x)\right)$ and $\boldsymbol{a(x)}=\left(a_1(x),\ldots,a_k(x)\right)$ are introduced with the canonical commutation relations:
\begin{eqnarray}\label{eq:ad1}
\nonumber
\left[a_i(x),a^\dag_j(x')\right]=\delta_{ij}\delta(x-x')I,\\
\left[a_i(x),a_j(x')\right]=\left[a^\dag_i(x),a^\dag_j(x')\right]=0,
\end{eqnarray}
where $x,x'\in\mathbf{R}^s$, $i,j=1,\ldots,k$. Normalized functional coherent states in the Fock space are defined as $|\boldsymbol{\xi}\rangle =\exp\left(-\frac{1}{2}\int d^sx|\boldsymbol{\xi}|^2\right)\exp\left(\int d^sx\boldsymbol{\xi}(x)\cdot\boldsymbol{a}^\dagger(x)\right)|\boldsymbol{0}\rangle$. They have the following property: \begin{equation}\label{eq:ad1a}
\boldsymbol{a}(x)|\boldsymbol{\xi}\rangle =\boldsymbol{\xi}(x)|\boldsymbol{\xi}\rangle,
\end{equation}.
Then the following vectors in the Fock space can be  introduced:
\begin{eqnarray}\label{eq:ad2}
|\xi,t\rangle = \exp\left[\frac{1}{2}\left(\int {d^s}x|\boldsymbol{\xi}|^2-\int {d^s}x|\boldsymbol{\xi}_0|^2\right)\right]|\boldsymbol{\xi}\rangle
=\exp\left(-\frac{1}{2}\int d^sx|\boldsymbol{\xi}_0|^2\right)
\times\exp\left(\int d^sx\boldsymbol{\xi}(x)\cdot\boldsymbol{a}^\dagger(x)\right)|\boldsymbol{0}\rangle.
\end{eqnarray}
Differentiation of equation ~(\ref{eq:ad2}) with respect to time $t$ yields, together with equation ~(\ref{eq:ad1a}), a linear Schr\"{o}dinger-like evolution equation in the Fock space:
\begin{eqnarray}\label{eq:ad3}
\frac{d}{dt}|\xi,t\rangle = M(t)|\xi,t\rangle,
|\xi,0\rangle=|\boldsymbol{\xi}_0\rangle,
\end{eqnarray}
where the boson "Hamiltonian" $M(t) = \int {d^s}x{\boldsymbol{a}^\dagger}(x)\cdot F(\boldsymbol{a}(x),{D^\alpha}\boldsymbol{a}(x))$.

Obviously, the majority of solutions of the linear equations in the Hilbert space have no predecessors among the solutions of the initial nonlinear equations i (3+1) - dimensional space-time, so the strict principle of superposition is abandoned; however, there is a "weak (or approximate) principle of superposition". Indeed, let us start with two different states in the Fock space corresponding (via the above procedure) to two different initial fields in 3 dimensions $\xi(t0,x)$ and $\psi(t0,x)$ (so these states are not the most general states in the Fock space). We can build a "weak superposition" of these states as follows: we build the following initial field in 3D: $a \xi+b \psi$, where $a$ and $b$ are the coefficients of the required superposition. Then we can build (using the above procedure) the state in the Fock space corresponding to $a \xi+b \psi$. If $\xi$ and $\psi$ are relatively weak, only vacuum state and a term linear in $\xi$ and $\psi$ will effectively survive in the expansion of the exponent for the coherent state. However, what we typically measure is the difference between the state and the vacuum state. So we have approximate superposition, at least at the initial moment. However, as electrodynamic interaction is rather weak (this is the basis of QED perturbation methods) and therefore nonlinearity of the evolution equations in 3D can be expected to be rather weak, this "superposition" will not differ much from the "true" superposition of the states in the Fock space. At least, one can expect this, until this issue is studied in detail.

\section{Bell Theorem}
In Section IV, it was shown that theories similar to quantum field theory (QFT) can be built that are basically equivalent to non-second-quantized scalar electrodynamics  and spinor electrodynamics on the set of solutions of the latter theories. However, the Bell inequalities cannot be violated in the local realistic theories, so this issue is discussed below using other people's arguments. Most of them were outlined by nightlight in various forums (see, e.g., Ref.~\cite{nightlight}) and by E. Santos (see, e.g., Ref.~\cite{Santos}), and can be summarized as follows.

While the Bell inequalities cannot be violated in local realistic theories, there are some reasons to believe these inequalities cannot be violated either in experiments or in quantum theory. Indeed, there seems to be  a consensus among experts that "a conclusive experiment falsifying in an absolutely uncontroversial way local realism is still missing"~\cite{Gen}. For example, A. Shimony offers the following opinion:

"The incompatibility of Local Realistic Theories with Quantum Mechanics permits adjudication by experiments, some of which are described here. Most of the dozens of experiments performed so far have favored Quantum Mechanics, but not decisively because of the "detection loophole" or the "communication loophole." The latter has been nearly decisively blocked by a recent experiment and there is a good prospect for blocking the former.~\cite{Shimony}"

M. Aspelmeyer and A. Zeilinger agree:

"But the ultimate test of Bell's theorem is still missing:
a single experiment that closes all the loopholes at once.
It is very unlikely that such an experiment will disagree
with the prediction of quantum mechanics, since this
would imply that nature makes use of both the detection
loophole in the Innsbruck experiment and of the
locality loophole in the NIST experiment. Nevertheless,
nature could be vicious, and such an experiment is desirable
if we are to finally close the book on local realism."~\cite{Aspel}

The popular argument of the latter quote that the loopholes were closed in separate experiments does not look conclusive either. Otherwise one could argue, for example, that the sum of the angles of a triangle in planar Euclidian geometry can differ from 180 degrees because experiments demonstrate that the sum of angles can differ from 180 degrees for planar quadrangles and for triangles on a sphere. The Bell inequalities for local realistic theories can only be guaranteed if all conditions of the Bell theorem are fulfilled simultaneously. Therefore, if one of the assumptions of the Bell theorem is not fulfilled in an experiment, the violation of the Bell inequalities  in that experiment cannot rule out local realistic theories.

On the other hand, to prove theoretically that the inequalities can be violated in quantum theory, one needs to use the projection postulate (loosely speaking, the postulate states that if some value of an observable is measured, the resulting state is an eigenstate of the relevant operator with the relevant eigenvalue). However, such postulate, strictly speaking, is in contradiction with the standard unitary evolution of the larger quantum system that includes the measured system and the measurement device (and the observer, if needed), as such postulate introduces irreversibility, whereas there is no irreversibility for the larger system (see, e.g., Ref.~\cite{alla} or the references to journal articles there), and, according to the quantum recurrence theorem, the larger system will return to a state that can be arbitrarily close to its initial, pre-measurement state, at least in a very large, but finite volume (Ref.~\cite{Bocc}).  Furthermore, unitary evolution cannot generate a mixture of states (the well-known measurement problem in quantum theory). The standard argument that collapse takes place during measurements and unitary evolution takes place between measurements does not seem convincing, as there is no obvious reason why unitary evolution should not be applicable to an instrument or an observer. For example, based on an analysis of experimental data, Schlosshauer (Ref.~\cite{Schloss}) believes that
 "(i) the universal validity of unitary dynamics and the superposition principle has been confirmed far into the mesoscopic and macroscopic realm in all experiments conducted thus far;
(ii) all observed ''restrictions'' can be correctly and completely accounted for by taking into account environmental decoherence effects;
(iii) no positive experimental evidence exists for physical state-vector
collapse;
(iv) the perception of single ''outcomes'' is likely to be explainable through decoherence effects in the neuronal apparatus."

 Therefore, it seems that mutually contradictory assumptions (e.g., unitary evolution and the projection postulate) are required to prove the Bell theorem, so it is on shaky grounds both theoretically and experimentally. On the other hand, the local realistic theories of this work reproduce unitary evolution of theories that look like quantum field theories, so they may need a modification of the theory of measurement (cf. Ref.~\cite{Santos2}).

\section{Conclusion}

Schr\"{o}dinger ~\cite{Schroed} noted that the complex charged matter field in the Klein-Gordon equation or in scalar electrodynamics can be made real by a gauge transform, although it is generally believed that complex functions are required to describe charged fields. Schr\"{o}dinger concludes his work (Ref.~\cite{Schroed}) with the following: "One is interested in what happens when [the Klein-Gordon equation] is replaced by Dirac's wave equation of 1927, or other first-order equations. This and the bearing on Dirac's 1951 theory will be discussed more fully elsewhere." To the best of this author's knowledge, Schr\"{o}dinger did not publish any sequel to Ref.~\cite{Schroed}, and the lack of extension to the Dirac equation may explain the fact that Schr\"{o}dinger's work did not get the attention it deserves. Such an extension is proposed here: it is shown that the Dirac equation is generally equivalent to one fourth-order partial differential equation for one complex component, which can also be made real by a gauge transform. Thus, the Dirac equation can be rewritten as an equation for one real function, rather than for a Dirac spinor. As the Dirac equation is one of the most fundamental, these results both belong in textbooks and can be used for development of new efficient methods and algorithms of quantum chemistry.

Furthermore, the matter field can be algebraically eliminated both from scalar electrodynamics (the Klein-Gordon-Maxwell equation) and from spinor electrodynamics (the Dirac-Maxwell electrodynamics) in a certain gauge (for spinor electrodynamics, this is done after introduction of a complex electromagnetic four-potential, which leaves the electromagnetic fields unchanged). The resulting equations describe independent dynamics of the electromagnetic field (they form closed systems of partial differential equations). It is also shown that for these systems of equations, a generalized Carleman linearization (Carleman embedding) procedure generates systems of linear equations in the Hilbert space, which look like second-quantized theories and are equivalent to the original nonlinear systems on the set of solutions of the latter. Thus, the relevant local realistic models can be embedded into quantum field theories. These models are generally equivalent to well-established models - scalar electrodynamics and spinor electrodynamics, so they correctly describe a large body of experimental data. Although they may need some modifications to achieve more complete agreement with experiments, they may be of great interest as "no drama quantum theories", as simple (in principle) as classical electrodynamics. Possible issues with the Bell theorem are discussed.

\section*{Acknowledgments}

The author is grateful to A.E. Allahverdyan, V.G. Bagrov, A.V. Gavrilin, A.Yu. Kamenshchik, A.Yu. Khrennikov, nightlight, H. Nikoli\textrm{$\mathrm{\acute{c}}$}, W. Struyve, R. Sverdlov, and H. Westman for their interest in this work and valuable remarks.

The author is also grateful to J. Noldus for useful discussions.

\end{document}